\documentclass[aps,preprint]{revtex4}%
\usepackage{amsfonts}
\usepackage{amsmath}
\usepackage{amssymb}
\usepackage{graphicx}
\usepackage{subfigure}
\usepackage{epstopdf}
\usepackage{float}
\usepackage{makecell}
\usepackage{booktabs}%
\allowdisplaybreaks
\makeatother
\begin{document}
\preprint{CTP-SCU/2020042}
\title{Thermodynamic Instability of $\mathrm{{3D}}$ Einstein-Born-Infeld AdS Black Holes}
\author{Hongmei Jing$^{a}$}
\email{hmjing@stu.scu.edu.cn}
\author{Benrong Mu$^{b}$}
\email{benrongmu@cdutcm.edu.cn}
\author{Jun Tao$^{a}$}
\email{taojun@scu.edu.cn, Corresponding author}
\author{Peng Wang$^{a}$}
\email{pengw@scu.edu.cn}
\affiliation{$^{a}$Center for Theoretical Physics, College of Physics, Sichuan University,
Chengdu, 610065, China}
\affiliation{$^{b}$Physics Teaching and Research Section, College of Medical Technology,
Chengdu University of Traditional Chinese Medicine, Chengdu 611137, China}

\begin{abstract}
Super-entropic black holes possess finite-area but noncompact event horizons
and violate the reverse isoperimetric inequality. It has been conjectured that
such black holes always have negative specific heat at constant volume $C_{V}$
or negative specific heat at constant pressure $C_{P}$ whenever $C_{V}>0$,
making them unstable in extended thermodynamics. In this paper, we test this
instability conjecture on a family of nonlinear electrodynamics black holes,
namely $3$D Einstein-Born-Infeld (EBI) AdS black holes. Our results show that
when nonlinear electrodynamics effects are weak, the instability conjecture is
valid. However, the conjecture can be violated in some parameter region when
nonlinear electrodynamics effects are strong enough. This observation thus
provides a counter example to the instability conjecture, which suggests that
super-entropic black holes could be thermodynamically stable.

\end{abstract}
\maketitle

\section{Introduction}

Black hole thermodynamics has been developed on the understanding of black
holes and quantum physics
\cite{Bekenstein:1973ur,Bekenstein:1974ax,Hawking:1974sw,Hawking:1976de,Grumiller:2014qma}%
. In the extended phase space, the cosmological constant can be considered as
a thermodynamic variable \cite{Kastor:2009wy,Wang:2006eb,Sekiwa:2006qj}.
Specifically, one can treat the cosmological constant $\Lambda$ as a
thermodynamic pressure $P=-\frac{\Lambda}{8\pi G}$, which gives that its
conjugate quantity is interpreted as a thermodynamic volume. For a negative
cosmological constant $\Lambda$, the pressure $P$ is positive, which yields a
well-defined equilibrium thermodynamic framework. It is worth noting that this
volume is usually not equal to the geometric volume of black holes except in
some simple cases, such as AdS Schwarzschild black holes.

Although the definition of thermodynamic volume has been given, its physical
interpretation is still a puzzle. An early attempt to answer this question
leads to the conjecture that the volume satisfies the reverse isoperimetric
inequality \cite{Cvetic:2010jb,Dolan:2013ft}. The conjecture was motivated in
the progress of studying Kerr-AdS black holes and then generalized to other
black holes. The reverse isoperimetric inequality is saturated for
Schwarzschild-AdS black holes, which indicates that for a black hole of a
given thermodynamic volume $V$, the entropy is maximized for Schwarzschild-AdS
black holes \cite{Cvetic:2010jb}. However, further investigations discovered
that this inequality does not apply to all kinds of black holes. Those black
holes that exceed the maximum entropy bound are called super-entropic black
holes
\cite{Hennigar:2014cfa,Hennigar:2015cja,Klemm:2014rda,Noorbakhsh:2017tbp,Noorbakhsh:2017nde}%
, which are rotating with non-compact event horizons of finite surface
area\textbf{ }\cite{Hennigar:2014cfa}. More relevant discussions can be found
in refs.
\cite{Appels:2019vow,Johnson:2019wcq,Imseis:2020vsw,Sinamuli:2015drn,Wu:2019xse,Wu:2020cgf,Frassino:2015oca,Johnson:2019mdp,Cong:2019bud,Feng:2017jub,Noda:2020vcn,Boudet:2020eyr,Xu:2020gzm,Wu:2020mby}%
.

A black hole reduces its mass-energy via the Hawking radiation. If the
specific heat is negative, this shrinking would lead to a higher temperature,
increased radiation, and hence more mass loss. Therefore, the system
accelerates through this downward spiral instead of settling into an
equilibrium state. For charged BTZ black holes, which are the simplest
super-entropic black holes, it has been shown that there is a connection
between the violation of the reverse isoperimetric inequality and the
thermodynamical instability with the specific heat at constant volume
$C_{V}<0$ \cite{Johnson:2019mdp}. This result consequently leads to a natural
conjecture that super-entropic black holes always have $C_{V}<0$, making them
unstable in the extended thermodynamics. Later, it was found that this
conjecture is violated for generalized exotic BTZ black holes in some
parameter region \cite{Cong:2019bud}. However in this case, the specific heat
at constant pressure $C_{P}$ was exhibited to be negative whenever $C_{V}>0$.
Thus, a broader version of the instability conjecture was proposed
\cite{Cong:2019bud}, which states that all super-entropic black holes are in
general thermodynamically unstable with either negative $C_{V}$ or negative
$C_{P}$. The instability conjecture was analytically verified for\textbf{ }%
$3$D charged BTZ black holes
\cite{Banados:1992wn,Banados:1992gq,Johnson:2019mdp}. Since it is difficult to
obtain analytical expressions for $C_{V}$ and $C_{P}$, using methods of ref.
\cite{Johnson:2019vqf}, the instability conjecture was numerically tested for
ultra spinning $d$-dimensional Kerr black holes \cite{Hennigar:2015cja},
generalized exotic BTZ black holes \cite{Cong:2019bud} and super-entropic
black hole with Immirzi in ref. \cite{Boudet:2020eyr}.

Here, we test the instability conjecture on EBI AdS black holes in
$(2+1)$-dimensional space-time. An EBI AdS black hole is the charged black
hole solution in the EBI theory based on the non-linear electrodynamics
proposed by Born and Infeld in 1934 \cite{Born:1934gh}, and is an extension of
a RN black hole in the Einstein-Maxwell theory. Since it was found that the
non-linear electrodynamics, in particular the Born-Infeld electrodynamics, can
come from the low-energy limit of string theory and encodes the low-energy
dynamics of D-branes (i.e., the low-energy effective action for a constant
electromagnetic field is precisely the Born-Infeld action)
\cite{Gibbons:2001gy,Fradkin:1985qd,Tseytlin:1986ti,Metsaev:1987qp}, it has
attracted considerable attention in recent years. After BI black hole
solutions in anti-de Sitter space were obtained \cite{Dey:2004yt,Cai:2004eh},
their properties have been extensively investigated
\cite{Cataldo:1999wr,Myung:2008kd,Fernando:2003tz,Hendi:2015hoa,Banerjee:2010da,Li:2016nll,Dehyadegari:2017hvd,Aiello:2004rz,Fernando:2006gh,Gunasekaran:2012dq,Zou:2013owa,Zeng:2016sei,Wang:2018xdz,Wang:2019kxp,Tao:2017fsy,Banerjee:2012vk,Ma:2020qkd,Bi:2020vcg}%
. Although various aspects of 3$\text{D}$ EBI AdS black holes have also been
studied, the specific heat and instability conjecture are expected to explore.

The organization of the rest of this work is as follows. In section
\ref{sec: thermodynamics of EBI black-hole}, we discuss thermodynamic
quantities of 3$\text{D}$ EBI AdS black holes. In section
\ref{sec:New-instability-conjecture}, we first show that 3$\text{D}$ EBI AdS
black holes violate the reverse isoperimetric inequality, and hence are
super-entropic. Then, the instability conjecture is considered by calculating
$C_{V}$ and $C_{P}$. We find that when non-linear electrodynamics effects are
strong enough, there exists some parameter region where $C_{V}$ and $C_{P}$
are both positive. This observation provides a counter example to the
instability conjecture. The conclusion is given in section
\ref{sec:Conclusion}. In this paper, we use geometrical units where $G$, $c$,
$\hbar$, and $k_{B}$ have been set to unity.

\section{Thermodynamics of $3\text{D}$ EBI AdS Black Holes}

\label{sec: thermodynamics of EBI black-hole}

In this section, thermodynamics of $\text{3D}$ EBI AdS black holes is
discussed. The action of the $3\text{D}$-Einstein gravity coupled with the
Born-Infeld electrodynamics is
\begin{align}
\label{eq:action}I  &  =\int d^{3}x\sqrt{-g}\left[  \frac{R-2\varLambda}%
{16\pi}+L(F)\right]  ,\\
L(F)  &  =\frac{b^{2}}{4\pi}\left(  1-\sqrt{1+\frac{2F}{b^{2}}}\right)  .
\end{align}
Here, the constant $b$ is the Born-Infeld parameter, $g$ is the determinant of
the metric tensor, $\Lambda=-1/l^{2}$ is the cosmological constant, $l$ is the
AdS radius, and $L(F)$ is the Lagrangian of the Born-Infeld electrodynamics.
The metric and gauge potential are \cite{Myung:2008kd,Cataldo:1999wr}
\begin{align}
ds^{2}  &  =-f\left(  r\right)  dt^{2}+f(r)^{-1}dr^{2}+r^{2}d\theta^{2},\\
f\left(  r\right)   &  =-8M+\frac{r^{2}}{l^{2}}+2b^{2}r\left(  r-\sqrt
{r^{2}+\frac{Q^{2}}{4b^{2}}}\right)  -\frac{1}{2}Q^{2}\ln\left[  r+\sqrt
{r^{2}+\frac{Q^{2}}{4b^{2}}}\right] \nonumber\\
&  +\frac{1}{2}Q^{2}\ln\left[  l+\sqrt{l^{2}+\frac{Q^{2}}{4b^{2}}}\right]
-2b^{2}l\left[  l-\sqrt{l^{2}+\frac{Q^{2}}{4b^{2}}}\right]  ,
\end{align}
where $Q$ and $M$ stand for the charge and mass of EBI black holes,
respectively. In the limit of $b\rightarrow\infty$, it reduces to the charged
BTZ black hole solution \cite{Johnson:2019mdp},
\[
f^{\text{BTZ}}\left(  r\right)  =-8M-\frac{Q^{2}}{2}\log\left(  \frac{r}%
{l}\right)  +\frac{r^{2}}{l^{2}}.
\]
The horizon is located at $r=r_{+}$ with $f\left(  r_{+}\right)  =0$, from
which the mass of $\text{3\text{D}}$ EBI AdS black holes is obtained
\cite{Myung:2008kd},
\begin{align}
M  &  =\frac{r_{+}^{2}}{8l^{2}}+\frac{1}{4}b^{2}r_{+}\left(  r_{+}-\sqrt
{r_{+}^{2}+\frac{Q^{2}}{4b^{2}}}\right)  -\frac{1}{16}Q^{2}\ln\left[
r_{+}+\sqrt{r_{+}^{2}+\frac{Q^{2}}{4b^{2}}}\right] \nonumber\\
&  +\frac{1}{16}Q^{2}\ln\left[  l+\sqrt{l^{2}+\frac{Q^{2}}{4b^{2}}}\right]
-\frac{1}{4}b^{2}l\left[  l-\sqrt{l^{2}+\frac{Q^{2}}{4b^{2}}}\right]  .
\label{mass}%
\end{align}
In the extended thermodynamics, one identifies the enthalpy $H$
\cite{Kastor:2009wy} with the mass of the black hole, and the pressure is
$P=-\Lambda/8\pi=1/8\pi l^{2}$. Moreover, the entropy $S$ is
\begin{equation}
S=\frac{A}{4}=\frac{1}{2}\pi r_{+}. \label{s}%
\end{equation}
The first law of thermodynamics, $dM=TdS+VdP+\Phi dQ$, gives the temperature
and the thermodynamic volume of $3\text{\text{D}}$ EBI AdS black holes%

\begin{align}
\label{volume-1}T  &  =\left.  \frac{\partial M}{\partial S}\right\vert
_{P}=\frac{r_{+}}{2\pi l^{2}}+\frac{b^{2}r_{+}}{\pi}\left(  1-\sqrt
{1+\frac{Q^{2}}{4b^{2}r_{+}^{2}} }\right)  ,\\
V  &  =\left.  \frac{\partial M}{\partial P}\right\vert _{S}=\pi r_{+}%
^{2}+2\pi l^{4}b^{2}\left(  1-\sqrt{1+\frac{Q^{2}}{4b^{2}l^{2}}}\right)  ,
\end{align}
respectively. It is observed that the thermodynamic volume is different from
the geometric volume $\pi r_{+}^{2}$.

\section{Instability Conjecture of $3\text{D}$ EBI AdS Black Holes}

\label{sec:New-instability-conjecture} For an asymptotically AdS black hole in
the extended phase space, it was conjectured in ref. \cite{Cvetic:2010jb} that
a reverse isoperimetric inequality holds,
\begin{equation}
\label{eq:RII}R\equiv\left(  \frac{\left(  d-1\right)  V}{\omega_{d-2}%
}\right)  ^{\frac{1}{d-1}}\left(  \frac{\omega_{d-2}}{A}\right)  ^{\frac
{1}{d-2}}\geq1,
\end{equation}
where the isoperimetric ratio $R$ is defined. Here, $V$ is the thermodynamic
volume, $A$ is the horizon area, $\omega_{d}$ stands for a $d$-dimensional
unit sphere,
\begin{equation}
\omega_{d}=\frac{2\pi^{\frac{d+1}{2}}}{\varGamma\left(  \frac{d+1}{2}\right)
},
\end{equation}
where $\omega_{1}=2\pi$ and $\omega_{2}=4\pi$. The reverse isoperimetric
inequality is saturated for a Schwarzschild AdS black hole since its
thermodynamic volume simply equals to its naive geometric volume. For some
more complicated black holes, e.g., Kerr \cite{Cvetic:2010jb}, STU
\cite{Caceres:2015vsa} and Taub-NUT/Bolt black holes \cite{Johnson:2014xza},
thermodynamic volumes are larger than naive geometric volumes, hence resulting
in $R>1$. Moreover unlike a Schwarzschild AdS black hole, these black holes
have nonzero $C_{V}$. However, several black hole solutions were later found
to violate the reverse isoperimetric inequality
\cite{Hennigar:2014cfa,Klemm:2014rda,Hennigar:2015cja,Brenna:2015pqa,Noorbakhsh:2016faj,Noorbakhsh:2017nde}%
. A black hole that violates the inequality is dubbed \textquotedblleft
super-entropic black hole\textquotedblright\ since its entropy is larger than
the maximum entropy allowed by the reverse isoperimetric inequality. Argued in
ref. \cite{Hennigar:2014cfa}, the violation is attributed to a result of the
finite-area but noncompact event horizon. It was further presented in refs.
\cite{Johnson:2019mdp,Cong:2019bud} that a large family of super-entropic
black holes has $C_{V}<0$ or $C_{P}<0$ whenever $C_{V}>0$, showing that they
are unstable in the extended thermodynamics.

In this section, we first show that $3\text{D}$ EBI AdS black holes are
super-entropic, which means that they violate the reverse isoperimetric
inequality $\left(  \ref{eq:RII}\right)  $. In fact, according to Eqs.
$\left(  \ref{s}\right)  $, $\left(  \ref{volume-1}\right)  $ and $\left(
\ref{eq:RII}\right)  $, the isoperimetric ratio $R$ for $3\text{\text{D}}$ EBI
AdS black holes can be readily computed to be
\begin{equation}
R=\sqrt{1-\frac{l^{2}Q^{2}}{2r_{+}^{2}}\left(  1+\sqrt{1+\frac{Q^{2}}%
{4l^{2}b^{2}}}\right)  ^{-1}}. \label{R}%
\end{equation}
It is obvious to observe from Eq. $\left(  \ref{R}\right)  $ that $R<1$, which
means $3\text{\text{D}}$ EBI AdS black holes violate the reverse isoperimetric
inequality as long as $Q\neq0$. Note that when $Q=0$, EBI AdS black holes
would reduce to Schwarzschild AdS black holes, which have $R=1$. Consequently,
$3\text{\text{D}}$ EBI AdS black holes are super-entropic. In the remainder of
this section, we discuss behavior of $C_{V}$ and $C_{P}$ of $3\text{\text{D}}$
EBI AdS black holes and provide further investigation for the instability conjecture.

Using Eq. $\left(  \ref{s}\right)  $, we can write thermodynamic quantities in
terms of $S$ and $P$,
\begin{align}
T  &  =\frac{8PS}{\pi}+\frac{2Sb^{2}}{\pi^{2}}\left(  1-\sqrt{1+\frac{\pi
^{2}Q^{2}}{16b^{2}S^{2}}}\right)  ,\label{T}\\
V  &  =\frac{4S^{2}}{\pi}+\frac{b^{2}}{32\pi P^{2}}\left(  1-\sqrt
{1+\frac{2\pi PQ^{2}}{b^{2}}}\right)  . \label{V}%
\end{align}
From Eq. $\left(  \ref{s}\right)  $, we observe that the entropy $S$ is
geometrical, and only depends on the horizon radius $r_{+}$. Hence the entropy
$S$ and the thermodynamic volume $V$ are independent functions, which
consequently gives a nonzero $C_{V}$. To obtain the specific heat at constant
volume $C_{V}$, it will be easier to start with $C_{P}$. Using Eq. $\left(
\ref{T}\right)  $, we can express $S$ in terms of $T$ and $P$,
\begin{equation}
S=\frac{\pi T}{16P}\left[  1+\left(  1+\frac{2\pi P}{b^{2}}\right)
^{-1}\left(  \frac{2\pi P}{b^{2}}+\sqrt{1+\frac{4P^{2}Q^{2}}{T^{2}b^{2}}%
+\frac{2PQ^{2}}{\pi T^{2}}}\right)  \right]  . \label{entropy}%
\end{equation}
Then $C_{P}\left(  T\right)  $ is given by
\begin{equation}
C_{P}(T)=\left.  T\frac{\partial S}{\partial T}\right\vert _{P}=\frac{\pi
T}{16P}\left[  1+\left(  1+\frac{2\pi P}{b^{2}}\right)  ^{-1}\left(
\frac{2\pi P}{b^{2}}+\frac{1}{\sqrt{1+\frac{4P^{2}Q^{2}}{b^{2}T^{2}}%
+\frac{2PQ^{2}}{\pi T^{2}}}}\right)  \right]  , \label{CP}%
\end{equation}
which is manifestly positive. For large $T$, one has $C_{P}(T)=\frac{\pi
T}{8P}+\cdots$. When $b\rightarrow\infty$, Eq. $\left(  \ref{CP}\right)  $
reduces to $C_{P}^{\text{BTZ}}$ of charged BTZ black holes (see Eq. $\left(
7\right)  $ in ref. \cite{Johnson:2019mdp}),%
\begin{equation}
C_{P}^{\text{BTZ}}(T)=\frac{\pi T}{16P}\left[  1+\frac{1}{\sqrt{1+\frac
{2PQ^{2}}{\pi T^{2}}}}\right]  .
\end{equation}

One can calculate $C_{V}(T)$ from $C_{P}(T)$ via the well-known relation,
\begin{equation}
\frac{C_{P}}{C_{V}}=\frac{1}{\kappa_{T}\beta_{S}}, \label{meyer formula}%
\end{equation}
where $\kappa_{T}\equiv-V\partial P/\left.  \partial V\right\vert _{T}$ is the
isothermal bulk modulus, and $\beta_{S}\equiv-V^{-1}\partial V/\left.
\partial P\right\vert _{S}$ is the adiabatic compressibility. To be
self-contained, a derivation of Eq. $\left(  \ref{meyer formula}\right)
$ is given in the appendix. Substituting Eq. $\left(  \ref{entropy}%
\right)  $ to Eq. $\left(  \ref{V}\right)  $ yields
\begin{align}
V\left(  T,P\right)   &  =\frac{\pi T^{2}}{64P^{2}}\left[  1+\left(
\frac{2\pi P}{b^{2}}+1\right)  ^{-1}\left(  \frac{2\pi P}{b^{2}}+\sqrt
{1+\frac{4P^{2}Q^{2}}{T^{2}b^{2}}+\frac{2PQ^{2}}{\pi T^{2}}}\right)  \right]
^{2}\nonumber\\
&  -\frac{Q^{2}}{16P}\left(  \sqrt{1+\frac{2\pi PQ^{2}}{b^{2}}}+1\right)
^{-1}. \label{volume}%
\end{align}
From Eqs. $\left(  \ref{V}\right)  $ and $\left(  \ref{volume}\right)  $,
$\kappa_{T}$ and $\beta_{S}$ can be readily computed,
\begin{align}
\kappa_{T}  &  \equiv-V\partial P/\left.  \partial V\right\vert _{T}=-V\left[
\frac{-8S^{2}}{\pi P}+\frac{TS}{2P}\gamma+\frac{1}{16P^{2}}\frac{Q^{2}}%
{\delta+1}+\frac{\pi Q^{4}}{16Pb^{2}}\frac{1}{\delta(\delta+1)^{2}}\right]
^{-1},\\
\beta_{S}  &  \equiv-V^{-1}\partial V/\left.  \partial P\right\vert
_{S}=-V^{-1}\left[  \frac{Q^{2}}{8P^{2}\left(  \sqrt{1+\frac{2\pi PQ^{2}%
}{b^{2}}}+1\right)  }-\frac{Q^{2}}{32P^{2}\sqrt{1+\frac{2\pi PQ^{2}}{b^{2}}}%
}\right]  ,
\end{align}
where
\begin{align}
\gamma &  =\frac{\frac{2\pi}{b^{2}}+\frac{\eta^{2}-1}{P\eta}-\frac{Q^{2}}%
{\eta\pi T^{2}}}{\left(  \frac{2\pi P}{b^{2}}+1\right)  }-\frac{\left(
\frac{2\pi P}{b^{2}}+\eta\right)  \left(  \frac{2\pi}{b^{2}}\right)  }{\left(
\frac{2\pi P}{b^{2}}+1\right)  ^{2}},\nonumber\\
\eta &  =\sqrt{1+\frac{4P^{2}Q^{2}}{T^{2}b^{2}}+\frac{2PQ^{2}}{\pi T^{2}}},\\
\delta &  =\sqrt{1+\frac{2\pi PQ^{2}}{b^{2}}}.\nonumber
\end{align}
With above results for $\kappa_{T}$, $\beta_{S}$ and $C_{P}(T)$, one can use
Eq. $\left(  \ref{meyer formula}\right)  $ to obtain the specific heat at
constant volume $C_{V}$. As a check, in the limit of $b\rightarrow\infty$, we
find that $C_{V}(T)$ becomes $C_{V}^{\text{BTZ}}(T)$ of charged BTZ black
holes (see Eq. $\left(  10\right)  $ in ref. \cite{Johnson:2019mdp}), where%
\begin{equation}
C_{V}^{\text{BTZ}}(T)=-\frac{Q^{2}}{32T}\left[  \frac{1+\sqrt{1+\frac{2PQ^{2}%
}{\pi T^{2}}}}{1+\sqrt{1+\frac{2PQ^{2}}{\pi T^{2}}}+\frac{3PQ^{2}}{2\pi T^{2}%
}}\right]  .
\end{equation}
\begin{figure}[ptb]
\centering
\subfigure[$C_{V}$ \& $C_{P}$ $vs$. $r_+$ and $M$ \& $T$ $vs$. $r_+$ for $b=100$.]{
		\includegraphics[width=0.45\linewidth]{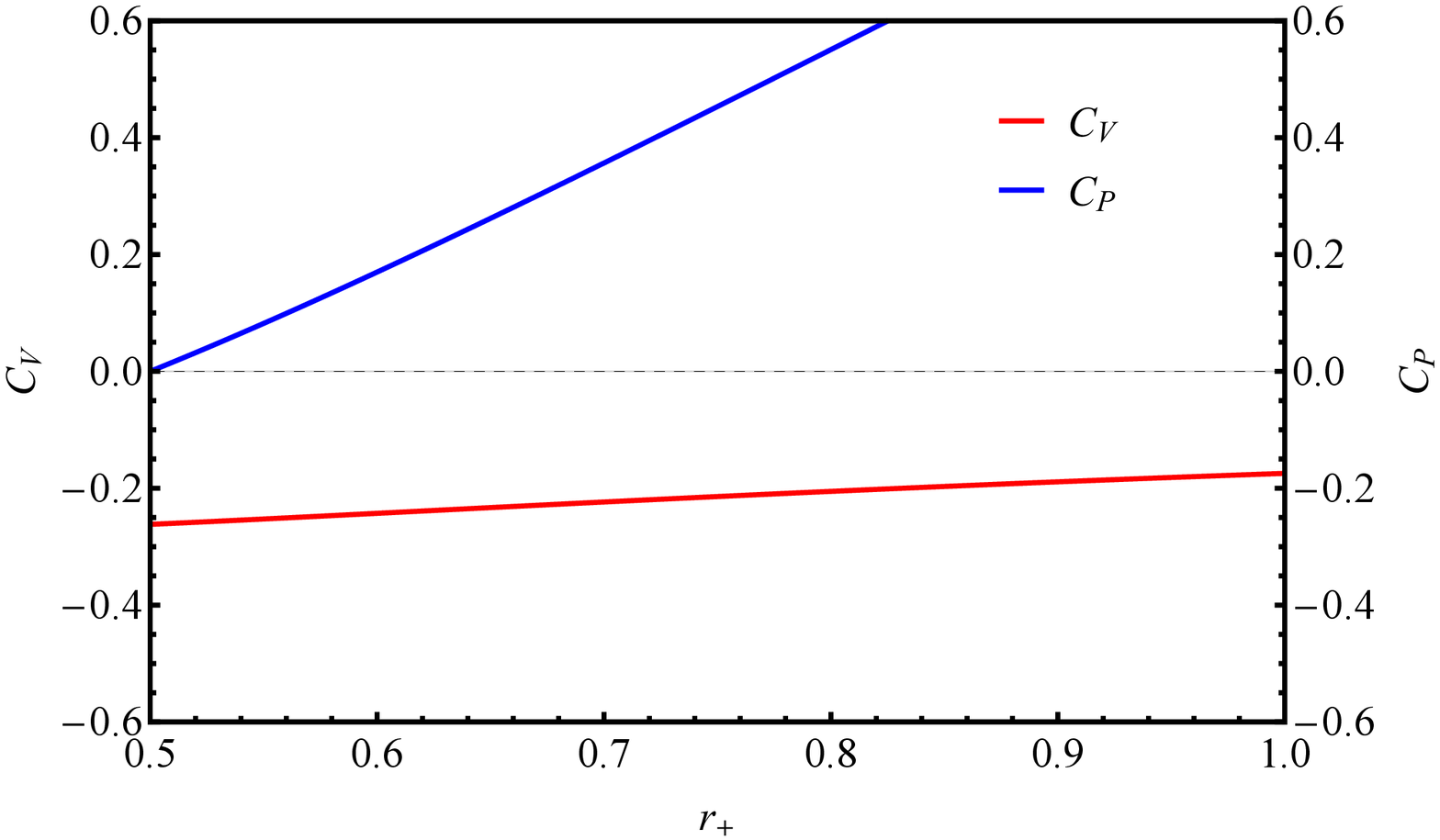}\hspace{2mm} \includegraphics[width=0.45\linewidth]{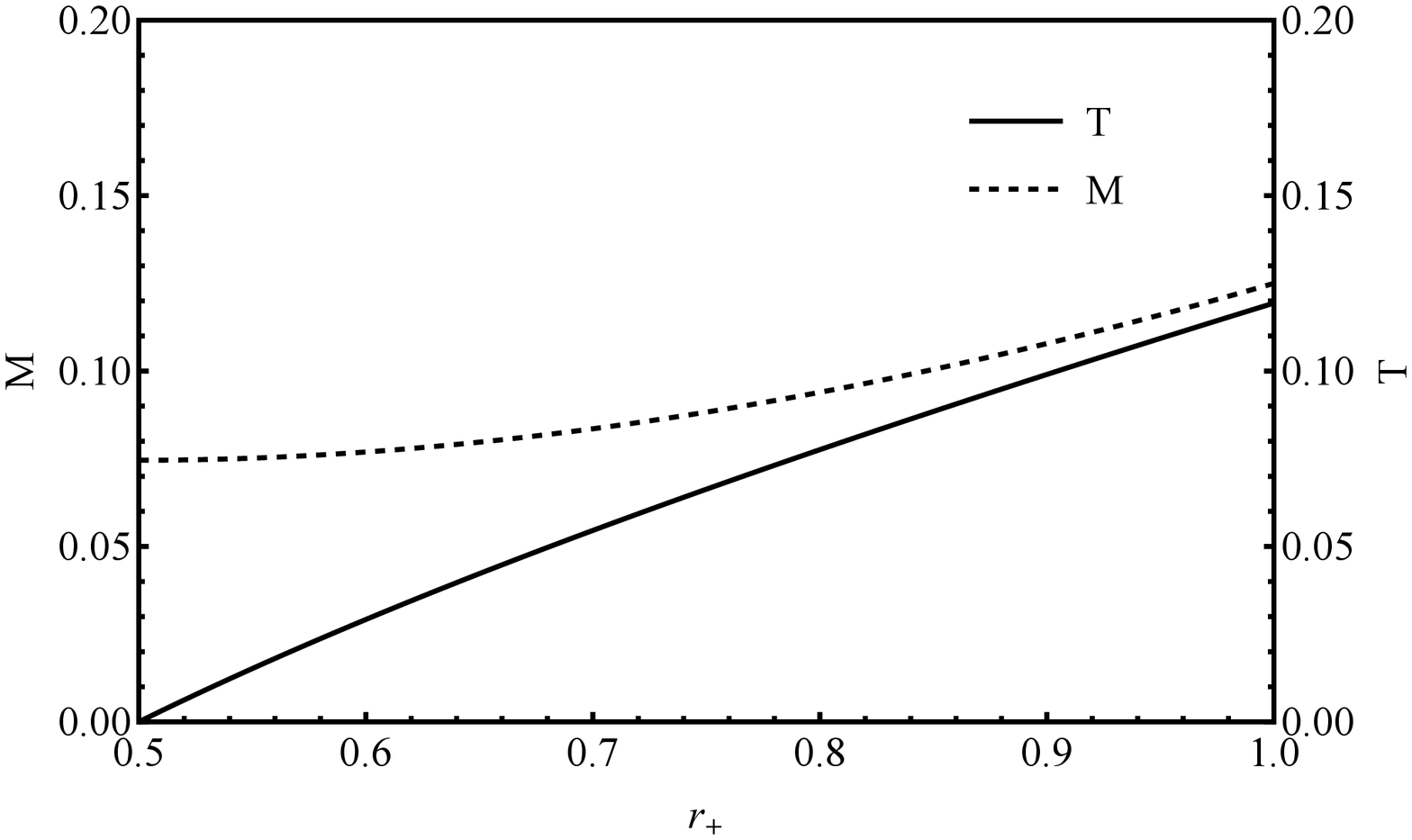}\label{Fig1:a}
	}
\subfigure[$C_{V}$ \& $C_{P}$ $vs$. $r_+$ and $M$ \& $T$ $vs$. $r_+$ for $b=1.69$.]{
		\includegraphics[width=0.45\linewidth]{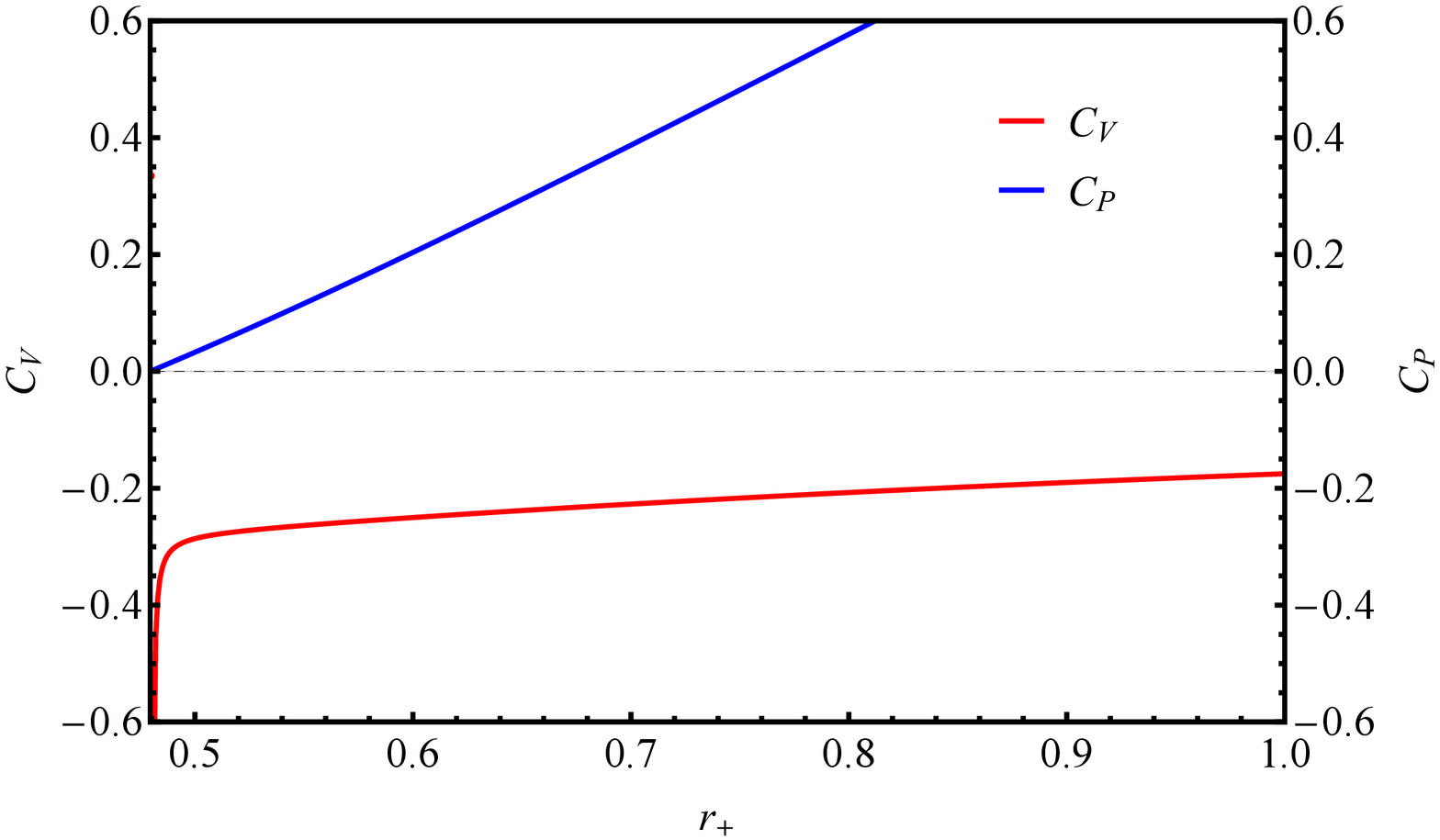}\hspace{2mm}
		\includegraphics[width=0.45\linewidth]{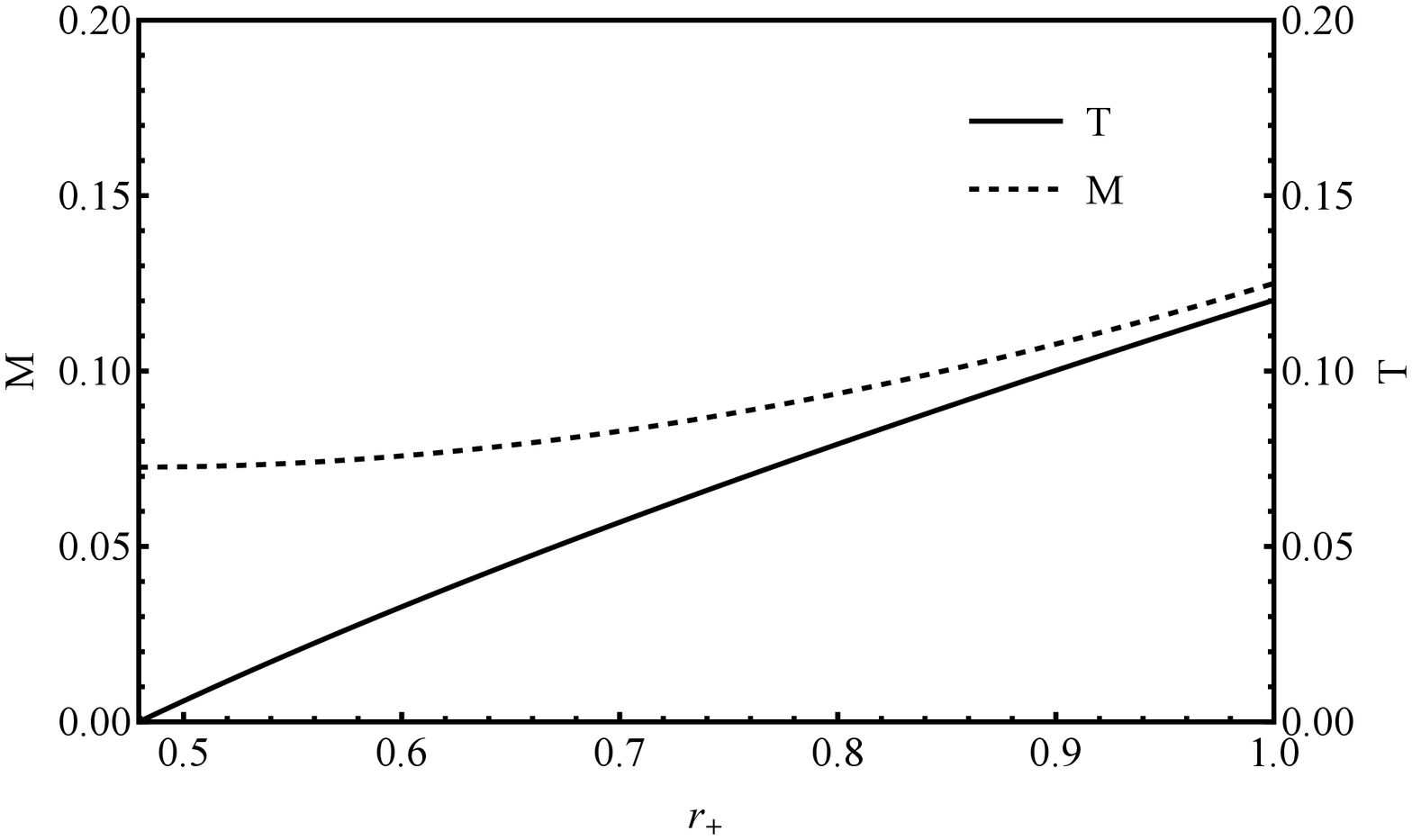}\label{Fig1:b}
	}
\subfigure[$C_{V}$ \& $C_{P}$ $vs$. $r_+$ and $M$ \& $T$ $vs$. $r_+$ for $b=0.5$.]{
		\includegraphics[width=0.45\linewidth]{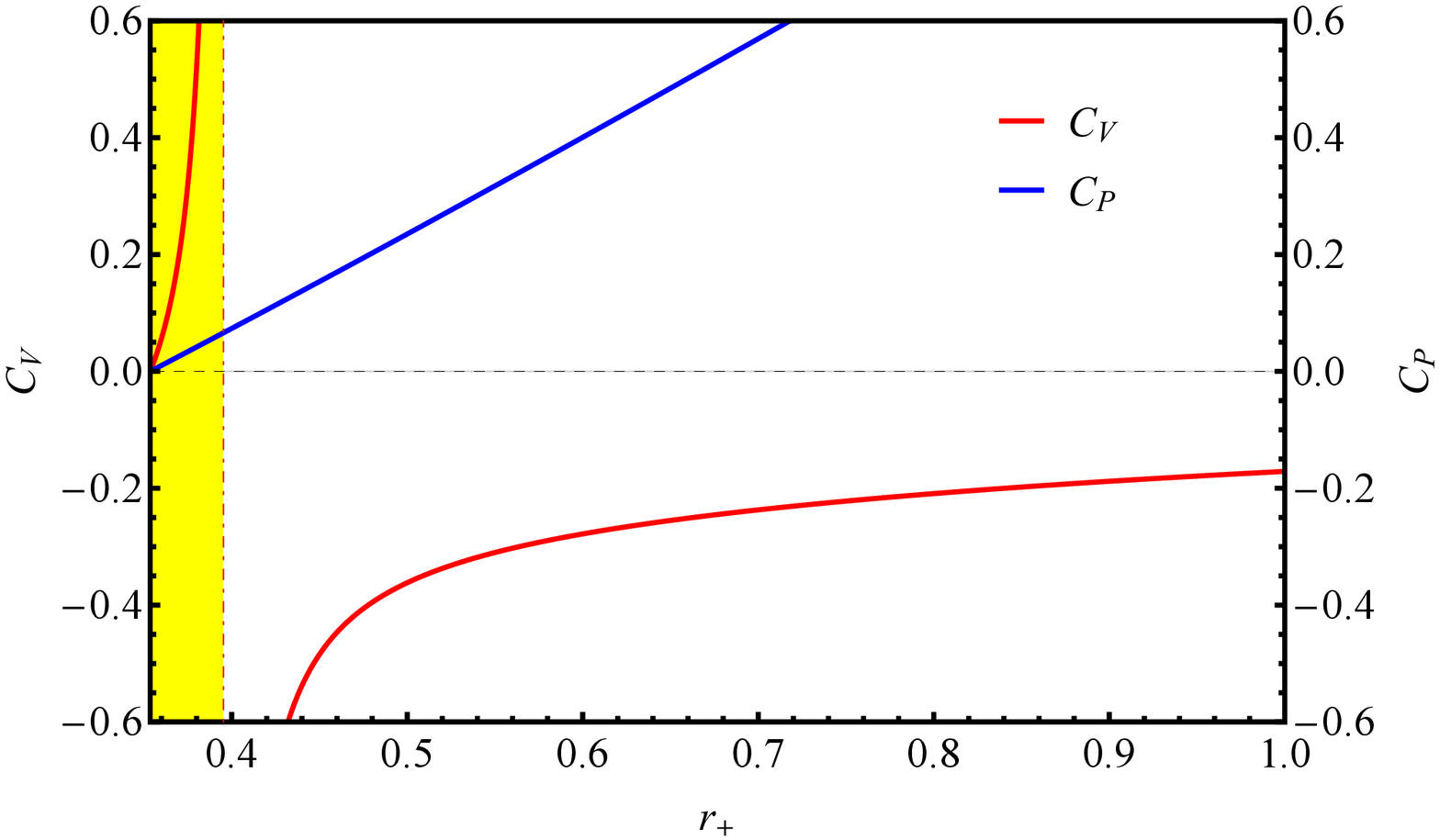}\hspace{2mm}
		\includegraphics[width=0.45\linewidth]{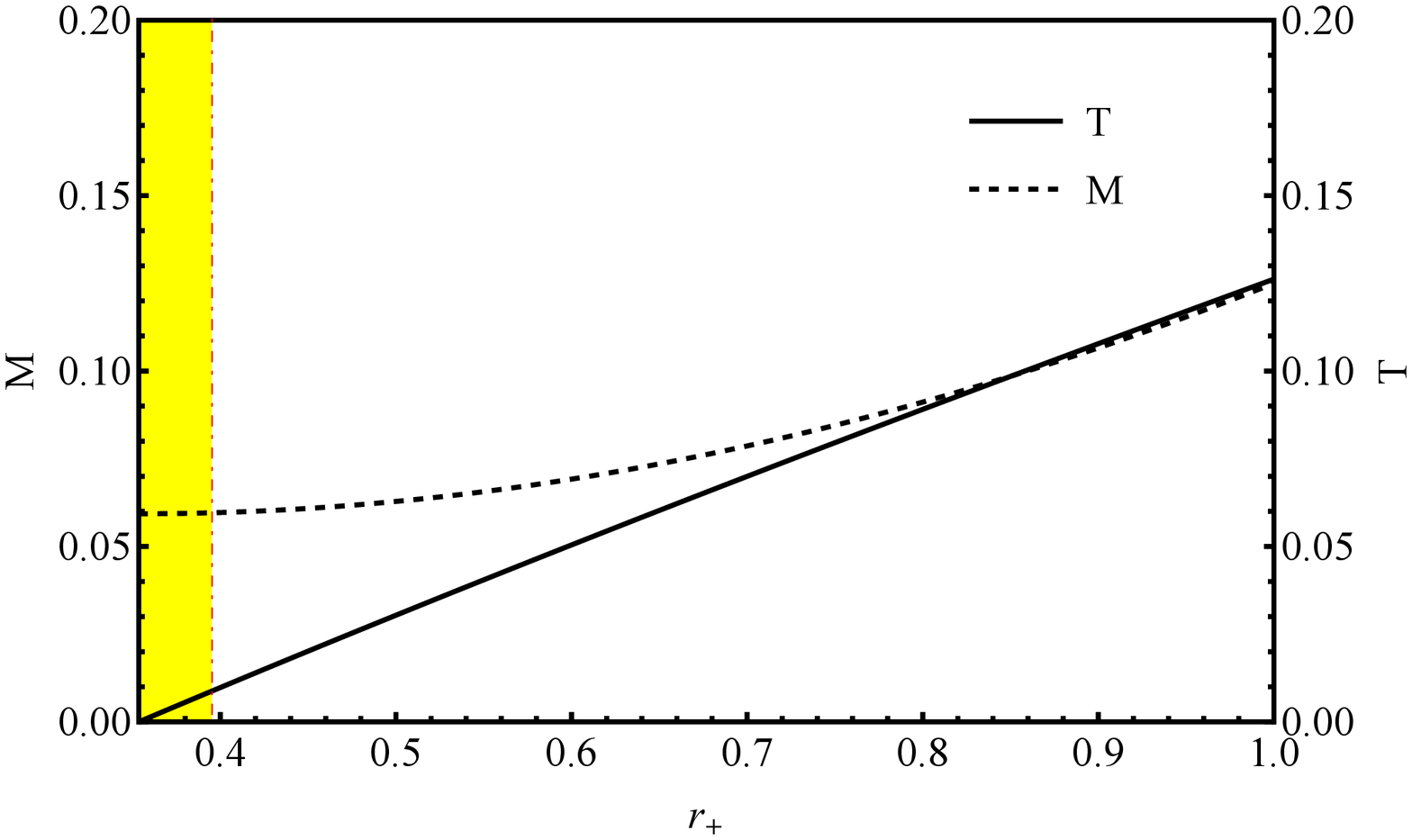}\label{Fig1:c}
	}
\subfigure[$C_{V}$ \& $C_{P}$ $vs$. $r_+$ and $M$ \& $T$ $vs$. $r_+$ for $b=0.1$.]{
		\includegraphics[width=0.45\linewidth]{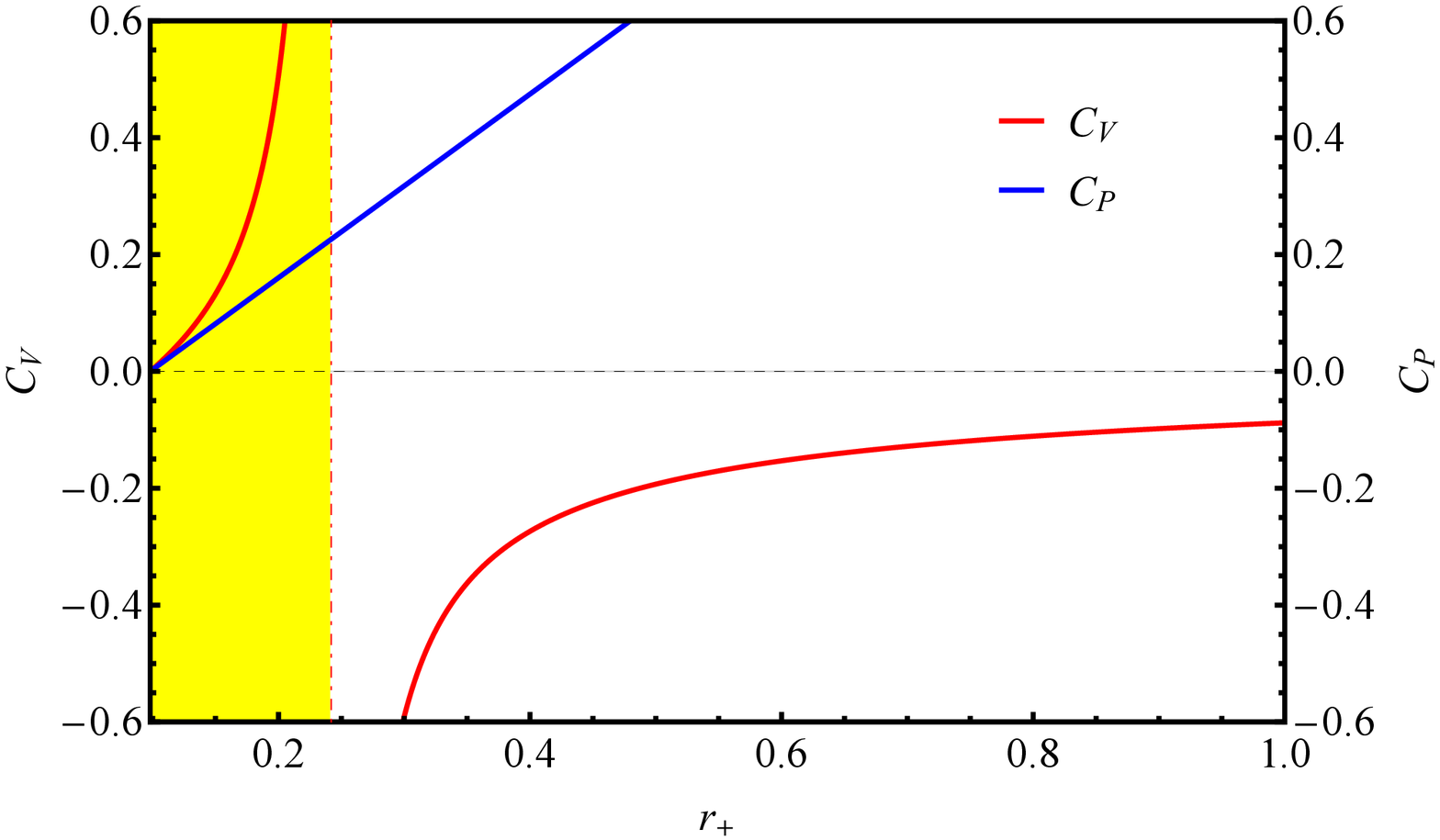}\hspace{2mm}
		\includegraphics[width=0.45\linewidth]{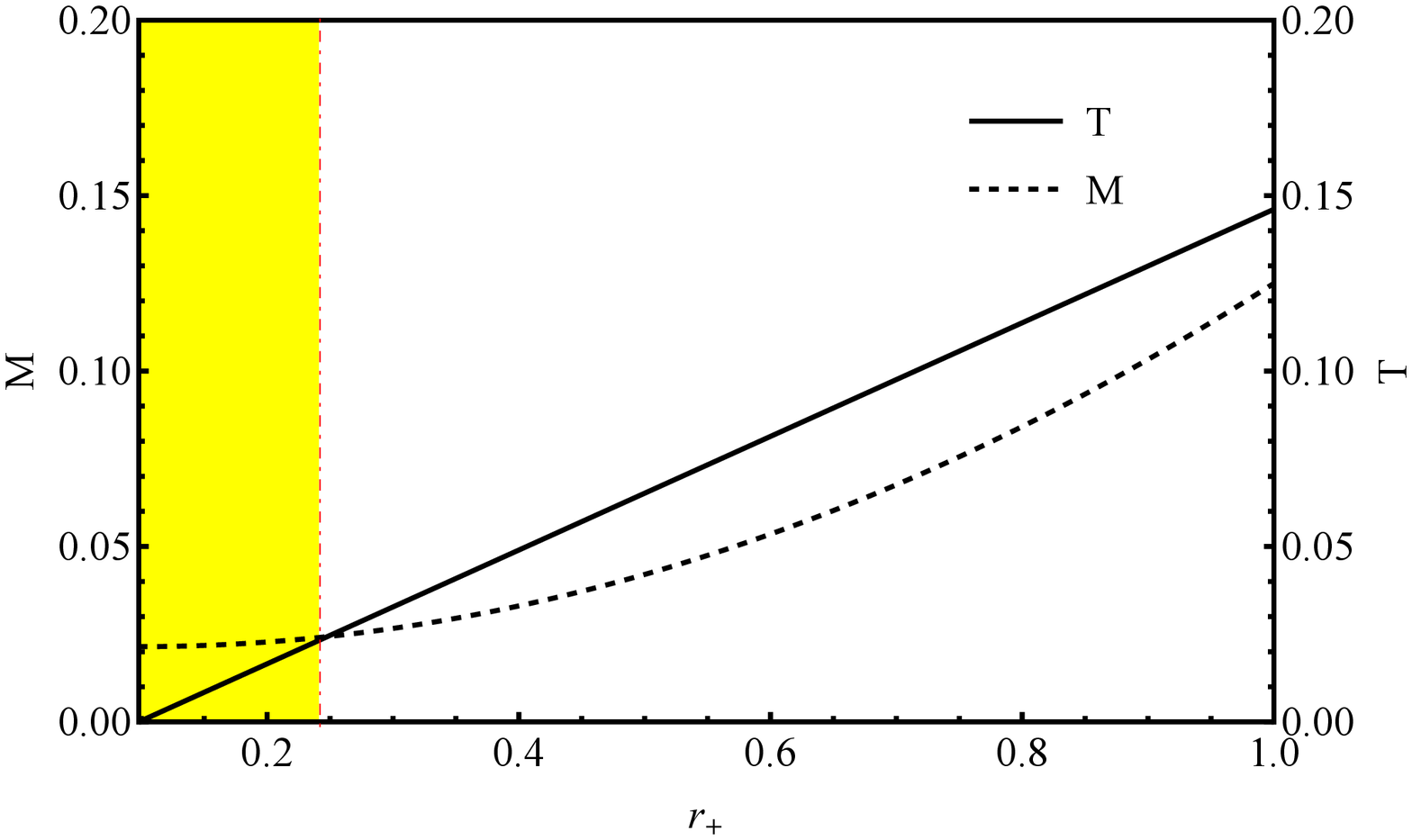}\label{Fig1:d}
	} \caption{{\footnotesize Plots of the heat capacity at constant volume
$C_{V}$, the heat capacity at constant pressure $C_{P}$, the black hole mass
$M$ and the black hole temperature $T$ against the black hole horizon radius
$r_{+}$ for $3\text{\text{D}}$ EBI AdS black holes with $Q=1=l$ and various
values of $b$. The yellow regions denote the regions of interest, where
$C_{V}$ and $C_{P}$ are both positive, and hence black holes can be free of
thermodynamic instability.}}%
\label{Fig1}%
\end{figure}

In FIG. \ref{Fig1}, we plot the specific heat at constant volume $C_{V}$, the
specific heat at constant pressure $C_{P}$, the black hole mass $M$ and the
black hole temperature $T$ as functions of the black hole horizon radius
$r_{+}$ for $3\text{\text{D}}$ EBI AdS black holes with fixed pressure
$l=1.0$, fixed charge $Q=1.0$ and various $b=0.1,0.5,1.69,100$. When $b=100$,
non-linear electrodynamics effects are negligible, and hence behavior of
$3\text{\text{D}}$ EBI AdS black holes closely resembles that of charged BTZ
black holes. As shown in FIG. \ref{Fig1:a}, $C_{P}$ is always positive whereas
$C_{V}$ is always negative, which recovers the results of BTZ black holes
\cite{Johnson:2019mdp}. As $b$ decreases to $b\simeq1.69$, FIG. \ref{Fig1:b}
exhibits that $C_{V}$ stays negative, and becomes large negative as $T$ goes
to zero. Interestingly, for small enough values of $b$ (i.e., $b\lesssim
1.69$), our numerical results show that $C_{V}$ and $C_{P}$ can both be
positive in some parameter region. In fact, when $b=0.5$ and $0.1$, the
regions where $C_{V}>0$ and $C_{P}>0$ are represented by yellow regions in
FIGs. \ref{Fig1:c} and \ref{Fig1:d}. Note that the black hole temperature $T$
and mass $M$ are both positive in the yellow regions of FIGs. \ref{Fig1:c} and
\ref{Fig1:d}, which means that the $3\text{\text{D}}$ EBI AdS black hole
solutions with $C_{V}>0$ and $C_{P}>0$ are physical. And FIGs. \ref{Fig1:c} \&
\ref{Fig1:d} suggest that the conjecture violation region increases in size
with decreasing parameter $b$. In short, we find that $3\text{D}$ EBI AdS
black holes can violate the instability conjecture.

\begin{figure}[ptb]
\centering
\subfigure{
		\includegraphics[width=0.45\linewidth]{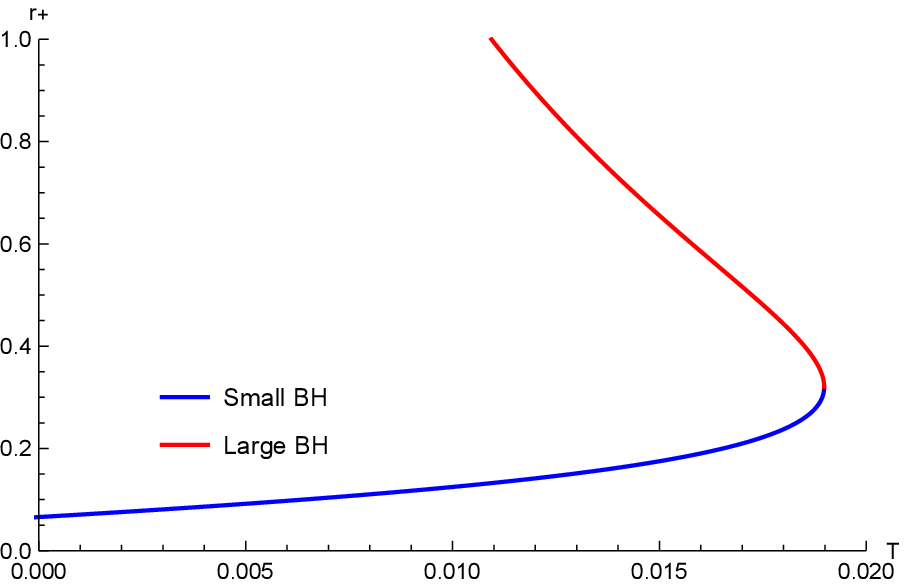}\hspace{2mm}
		\includegraphics[width=0.45\linewidth]{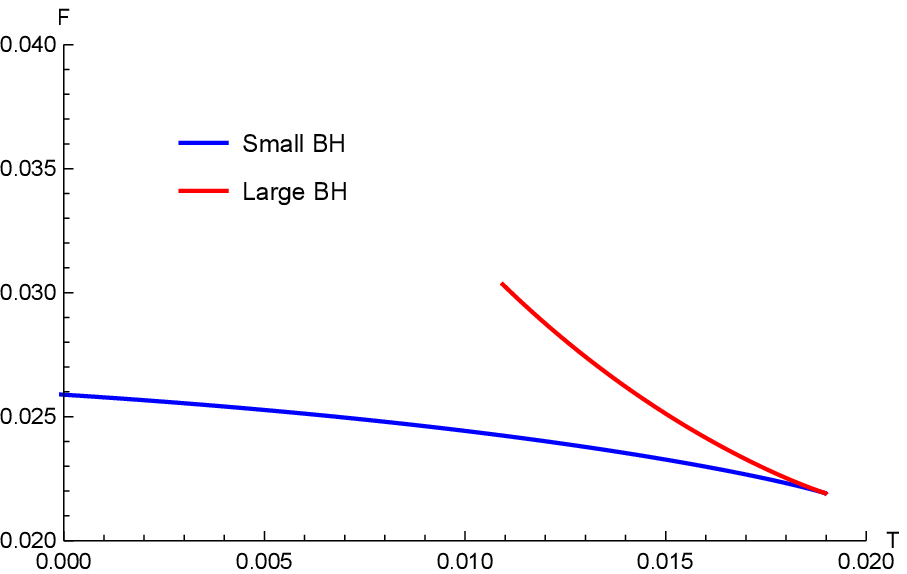}
	}\caption{\footnotesize Plots of the horizon radius $r_{+}$ and the Helmholtz free energy
$F$ against the black hole temperature $T$ for $3\text{\text{D}}$ EBI AdS
black holes with fixed volume $V$. Here, we take $Q=1$ and $b=0.1$. The blue
and red lines represent Small BH and Large BH, respectively. The specific heat
at constant volume $C_{V}$ of Small/Large BH is positive/negative. As a
result, $C_{V}$ is discontinuous at the maximum value of $T$.}%
\label{Fig2}%
\end{figure}

Interestingly, FIG. \ref{Fig1} shows that $C_{V}$ has a discontinuity for a
small enough $b$. To investigate the nature of the discontinuity of $C_{V}$,
we plot the horizon radius $r_{+}$ and the Helmholtz free energy $F$ as
functions of the black hole temperature $T$ with fixed volume $V$ in FIG.
\ref{Fig2}, where $Q=1$ and $b=0.1$. The left panel of FIG. \ref{Fig2} shows
that, for a given $T$, there are two black hole solutions of different sizes,
namely Large BH (red line) and Small BH (blue line). Moreover, the black hole
temperature $T$ has a maximum $T_{\max}$, which corresponds to $\left.
\partial r_{+}/\partial T\right\vert _{V}=0$. Note that $C_{V}$ can be
rewritten as%
\begin{equation}
C_{V}=\frac{\pi}{2}\left.  \frac{\partial r_{+}}{\partial T}\right\vert _{V},
\end{equation}
where we use Eq. $\left(  \ref{s}\right)  $ for the entropy $S$. Therefore,
Large/Small BH has a negative/positive $C_{V}$, which goes to
negative/positive infinity as $T$ approaches $T_{\max}$. In short, the
discontinuity of $C_{V}$ corresponds to the maximum value of the black hole
temperature, where the two black hole phases (i.e., Large BH and Small BH) merge.
The right panel of FIG. \ref{Fig2} displays that the free energy of Small BH
is always smaller than that of Large BH, which indicates that there is no
phase transition. Our results suggest that, at a constant volume,
$3\text{\text{D}}$ EBI AdS black holes with positive $C_{V}$ are globally
stable.

\section{Conclusion and discussion}

\label{sec:Conclusion}

In this paper, considering $3\text{D}$ EBI AdS black holes, we tested the
instability conjecture: super-entropic black holes always have $C_{V}<0$ or
$C_{P}<0$ whenever $C_{V}>0$, making them unstable in extended gravitational
thermodynamics. This conjecture was tested and found satisfied for a large
class of super-entropic solutions
\cite{Johnson:2019mdp,Hennigar:2014cfa,Cong:2019bud}. After showing
$3\text{D}$ EBI AdS black holes are super-entropic, we found that the black
holes satisfy the instability conjecture when $b$ is large enough (i.e.,
non-linear electrodynamics effects are inessential). However, when non-linear
electrodynamics effects play an important role, our numerical results (see
FIG. \ref{Fig1}) showed that there exists some parameter region, where both
$C_{V}>0$ and $C_{P}>0$, and hence provided a counter example to the
instability conjecture. In addition, it was suggested that the violation
region will increase as non-linear electrodynamics effects become stronger.

It is worthwhile pointing out that for $d\geq4$ dimension, the thermodynamic
volume of EBI AdS black holes is just the naive geometric volume $V=\left(
d-1\right)  ^{-1}\omega_{d-2}r_{+}^{d-1}$, which means that $R=1$, and hence
higher dimensional EBI AdS black holes are not super-entropic. On the other
hand, for higher dimensional EBI AdS black holes, the entropy $S$ and the
volume $V$ are not independent, which leads to the constant volume specific
heat $C_{V}=0$ \cite{Dolan:2010ha}.

\begin{acknowledgments}
We are grateful to Wei Hong and Yucheng Huang for useful discussions and
numerical analysis. This work is supported in part by NSFC (Grant No.
11005016, 11875196 and 11375121).
\end{acknowledgments}

\appendix

\section{Derivation of Eq. $\left(  \ref{meyer formula}\right)  $}

In this appendix, we present a derivation of Eq. $\left(
\ref{meyer formula}\right)  $, which starts from the definition of the
adiabatic compressibility and the isothermal bulk modulus. In fact, the
adiabatic compressibility $\beta_{S}$ and the isothermal bulk modulus
$\kappa_{T}$ are defined by
\begin{equation}
\beta_{S}\equiv-\frac{1}{V}\left.  \frac{\partial V}{\partial P}\right\vert
_{S}\text{ and }\kappa_{T}\equiv-V\left.  \frac{\partial P}{\partial
V}\right\vert _{T},
\end{equation}
respectively. Using properties of partial derivatives, one can rewrite
$\beta_{S}$ as
\begin{equation}
\beta_{S}=-\frac{1}{V}\left.  \frac{\partial V}{\partial S}\right\vert
_{P}\left.  \frac{\partial S}{\partial P}\right\vert _{V}=-\frac{1}{V}%
\frac{\left.  \frac{\partial S}{\partial P}\right\vert _{V}}{\left.
\frac{\partial S}{\partial V}\right\vert _{P}}=-\frac{1}{V}\frac{\left.
\frac{\partial S}{\partial T}\right\vert _{V}\left.  \frac{\partial
T}{\partial P}\right\vert _{V}}{\left.  \frac{\partial S}{\partial
T}\right\vert _{P}\left.  \frac{\partial T}{\partial V}\right\vert _{P}%
}.\label{the adiabatic compressibility}%
\end{equation}
On the other hand, the specific heat at constant pressure $C_{P}$ and the
specific heat at constant volume $C_{V}$ are defined by
\begin{equation}
C_{P}\equiv T\left.  \frac{\partial S}{\partial T}\right\vert _{P}\text{ and
}C_{V}\equiv T\left.  \frac{\partial S}{\partial T}\right\vert _{V},
\end{equation}
respectively. Consequently, Eq. $\left(  \ref{the adiabatic compressibility}%
\right)  $ reduces to
\begin{equation}
\beta_{S}=\frac{C_{V}}{C_{P}}\left(  -\frac{1}{V}\left.  \frac{\partial
T}{\partial P}\right\vert _{V}\left.  \frac{\partial V}{\partial T}\right\vert
_{P}\right)  =\frac{C_{V}}{C_{P}}\left(  -\frac{1}{V}\left.  \frac{\partial
V}{\partial P}\right\vert _{T}\right)  =\frac{C_{V}}{C_{P}}\frac{1}{\kappa
_{T}},
\end{equation}
which leads to
\begin{equation}
\frac{C_{P}}{C_{V}}=\frac{1}{\beta_{S}\kappa_{T}}.
\end{equation}

\end{document}